\documentclass[fleqn,twoside]{article}
\usepackage{espcrc2}

\readRCS
$Id: espcrc2.tex,v 1.2 2004/02/24 11:22:11 spepping Exp $
\usepackage{graphicx}
\usepackage[figuresright]{rotating}

\newcommand{\fr}[1]{
             \frac{#1}}
\newcommand{\bea}{\begin{eqnarray}}
\newcommand{\eea}{\end{eqnarray}}

\newcommand{\chibar}{\overline{\chi}}

\newcommand{\ket}{{\rangle }}

\newcommand{\bra}{{\langle }}
\newcommand{\gc}{\bra\fr{\alpha_s}{\pi}G^2\ket}

\newcommand{\ga}{{g_{{\mathcal A}}}}

\newcommand{\AmS}{{\protect\the\textfont2
  A\kern-.1667em\lower.5ex\hbox{M}\kern-.125emS}}

\title{Non-factorizable contributions to
$\overline{B^0_d} \rightarrow D_s^{(*)} \overline{D_s^{(*)}}$ }

\author{J.O. Eeg\address{Department of Physics, University of Oslo, P.O. Box 1048
 Blindern, N-0316 Oslo, Norway}
\thanks{Presented by J.O.Eeg. Supported in part by the Norwegian
 research council
 and  by the European Union RTN
network, Contract No. HPRN-CT-2002-00311  (EURIDICE).},
S. Fajfer\address[IJS]{J. Stefan Institute, Jamova 39,  SI-1000 Ljubljana, 
Slovenia}\address{Physics Department, University of Ljubljana , Jadranska 19,
SI-1000 Ljubljana, Slovenia}
\thanks{Supported in part
 by the Ministry of Education,
Science and Sport of the Republic of Slovenia.},
A. Hiorth\address{RF-Rogaland Research, P.O.Box 8046, N-4068 Stavanger, Norway}
and A. Prapotnik\addressmark[IJS]}

\begin{document}

\begin{abstract}
It is pointed out that decays of the type $B \rightarrow D
\overline{D}$
have no factorizable contributions, unless at least one of the charmed mesons
in the final state is a vector meson. 
The dominant contributions to the decay amplitudes 
arise from  chiral loop contributions and 
tree level amplitudes generated by soft gluon emissions
forming a gluon condensate.
We predict that the 
branching ratios   for the processes $\bar B^0 \to D_s^+ D_s^-$, 
  $\bar B^0 \to D_s^{+*} D_s^- $ and 
 $\bar B^0 \to D_s^+ D_s^{-*}$ are all of order $(3- 4) \times 10^{-4}$,
while  $\bar B^0 \to D_s^{+*} D_s^{-*}$ has a branching ratio 5 to 10
times bigger. We emphasize that the branching ratios are sensitive to 
$1/m_c$ corrections.

\vspace{1pc}
\end{abstract}

\maketitle

\section{INTRODUCTION}

Within the standard 
approach  for non-leptonic decays  
one constructs an effective Lagrangian  in terms of
(mainly) four quark operators multiplied with Wilson coefficients.
The simple and naive assumption that the matrix elements of the four
quark operators factorize as  the product of two current matrix elements
can be shown to be valid in the strict limit $N_c \rightarrow \infty$,
where $N_c$ is the number of colors. 
 In the treatment of $B$-meson 
decays with energy release of the order of the $b$-quark mass,  usually the 
factorization assumption or  QCD factorization has been used. 
However, for decays where the energy release is of order 1~GeV, 
QCD factorization is not expected to hold. 
Here we discuss non-factorizable contributions to 
 the decay modes  $\overline{B^0_d} \rightarrow D_s^{(*)}
 \overline{D_s^{(*)}}$, where $ D_s^{(*)}$ is a pseudoscalar or a
 vector meson.
At quark level
such decays occur through the 
annihilation mechanism
$b \bar{q} \rightarrow c \bar{c}$, where $q=d,s$ 
respectively. However, within the factorized limit the annihilation
 mechanism will give a vanishing amplitude 
 due to current conservation, unless one or two of the $D$-mesons in
 the final state are vectors. These 
 contributions are proportional to a numerically non-favored  Wilson coefficient.

In contrast, 
 the typical   factorized   decay modes  which proceed
through the spectator mechanism, say
$\overline{B^0} \rightarrow D^+ D_s^-$, 
are proportional to the numerically favored   Wilson 
coefficient.
In our approach  \cite{EFH}, the non-factorizable  
contributions are coming from  the 
chiral loops and from tree level amplitudes generated by soft gluon
 emision forming a  gluon condensate.
The  gluon condensate contributions
can be 
calculated within 
a recently developed Heavy Light Chiral Quark Model (HL$\chi$QM) \cite{ahjoe}.
This model has been applied to processes involving $B$-mesons in 
\cite{ahjoe,ahjoeB}.
Both the chiral loop contributions and the gluon condensate
 contributions are 
 proportional to the numerically favorable Wilson coefficient.

\section{FRAMEWORK}

\subsection{Effective Lagrangian at quark level}

The effective Lagrangian at quark level reads:
\begin{equation} {\mathcal L}_{W}=  - \frac{G_F}{\sqrt{2}} V_{cb}V_{cq}^*
\sum_i a_i(\mu) \; Q_i (\mu) \,,
\label{Lquark}
\end{equation}
where $q=d,s$ and $a_i(\mu)$ are  Wilson coefficients that
carry all information of the short distance  physics above the
renormalization scale $\mu$.
 The matrix elements of $Q_i(\mu)$ 
contain all non-perturbative, long distance  physics below
$\mu$.
Within Heavy Quark Effective Theory~(HQEFT)
the effective non-leptonic 
Lagrangian ${\mathcal L}_{W}$ can be  evolved down to the scale
 $\mu = \Lambda_\chi \simeq$ 1 GeV, and below $\mu=m_c$ the Wilson
coefficients $a_i$ are complex \cite{GKMWF}. 

 The  numerically relevant operators in our case are 
\begin{eqnarray}
Q_{1}= 4(\overline{q}_L \gamma^\mu b_L) \; ( \overline{c}_L
\gamma_\mu c_L )\,,
\nonumber \\
Q_{2}= 4( \overline{c}_L \gamma^\mu b_L ) \; ( \overline{q}_L
\gamma_\mu c_L ) \,, 
\label{Q12} 
\end{eqnarray}  
where  $L$
denotes a left-handed particle.
At $\mu =
\Lambda_\chi$,  which by construction is the matching
scale within our approach\cite{EFH,ahjoe,ahjoeB}, 
one finds $a_1 \simeq -0.35 -0.07i$
and $a_2 \simeq 1.29+ 0.08i$.
 In the next section we will see how
 the currents in the operators in (\ref{Q12}) are bosonized.

In order to obtain all matrix elements of
the Lagrangian (\ref{Lquark}) we need the 
 Fierz transformed version of the operators in  (\ref{Q12}).
To find  these, we use the relation:
\begin{equation}
\delta_{i j}\delta_{l n}  =   \frac{1}{N_c} \delta_{i n} \delta_{l j}
 \; +  \; 2 \; t_{i n}^a \; t_{l j}^a \, ,
\label{fierz}
\end{equation}
where $i$, $j$, $k$ and $n$ are color indices running from 1 to 3 and
$a$ is a color octet index. One obtains
\begin{equation}
Q_1^F = \frac{1}{N_C} Q_2 + \widetilde Q_2 \; ,
  \quad Q_2^F = \frac{1}{N_C} Q_1 + \widetilde Q_1 \; ,
\label{Q12F}
\end{equation}
where the superscript $F$ means ``Fierzed'', and 
\begin{eqnarray}
\widetilde{Q_{1}}  = 4  (\overline{q}_L \gamma^\mu t^a  b_L )  \; \,
           ( \overline{c}_L \gamma_\mu t^a c_L ) \,,
\nonumber \\
\widetilde{Q_{2}}  =  4 \,  ( \overline{c}_L \gamma^\mu t^a b_L )  \; \,
           ( \overline{q}_L \gamma_\mu t^a c_L ) \,,
\label{QCol} 
\end{eqnarray}  
where $t^a$ denotes the color matrices. These expressions are
used to obtain the gluon condensate contributions. 

\subsection{Heavy light chiral perturbation theory}

The Heavy Quark
Effective Theory (HQEFT) 
 Lagrangian is:
\begin{equation}
{\mathcal L}_{HQEFT} = \pm \overline{Q_v^{(\pm)}} \, i v \cdot D \,
Q_v^{(\pm)} + {\mathcal O}(m_Q^{- 1}) \; ,
\label{LHQEFT}
\end{equation}
where 
 $Q_v^{(+)}(x)$ is a (reduced)  heavy quark field ($b$ or $c$ in our
 case) with velocity $v$, and  $Q_v^{(-)}(x)$ is the field of a heavy
 anti-quark ($\bar{c}$ in our case). Furthermore, $m_Q$ is the heavy
 quark mass, and
 $D_\mu$ is the covariant derivative containing the gluon field.

After integrating out the heavy and light quarks, the effective Lagrangian
up to ${\mathcal O}(m_Q^{-1})$ can be written as a kinetic term plus a
term
describing the chiral interaction between heavy and light mesons
\cite{ahjoe}:
\begin{eqnarray}
{\mathcal L}_\chi  = - \ga \, Tr\left[\overline{H^{(\pm)}_{a}}H^{(\pm)}_{b}
\gamma_\mu\gamma_5 {\mathcal A}^\mu_{ba}\right]\, 
,\label{LS1}
\end{eqnarray}
where $H_a^{(\pm)}$ is the heavy meson field containing a spin zero
and a spin one boson:
\begin{eqnarray}
 H_a^{(\pm)}   =   P_{\pm} (P_{a \mu}^{(\pm)} \gamma^\mu - i P_{a
5}^{(\pm)} \gamma_5) \; \; .
\label{barH}
\end{eqnarray}
Here $a,b$ are
flavor indices and  $P_\pm=(1 \pm \gamma \cdot v)/2$ are projecting operators.  
 The  axial vector field  ${\mathcal A}_{\mu}$ in (\ref{LS1}) is defined as:
\begin{equation}
 {\mathcal A}_\mu =  -
\frac{i}{2} (\xi^\dagger\partial_\mu\xi -\xi\partial_\mu\xi^\dagger)\,, 
\label{defVA}
\end{equation}
where $ \xi\equiv exp[i(\Pi/f)]$. Moreover,  $f$ is the bare pion
 coupling
 and $\Pi$ is a 3 by 3 matrix
which contains the Goldstone bosons $\pi,K,\eta$ in the standard way.
The axial chiral coupling $\ga$ is $\simeq 0.6$.

\begin{figure}[t]
\begin{center}
\includegraphics[width=6cm]{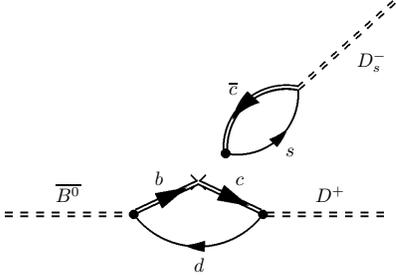}
\caption{\small{Factorized contribution for 
$\overline{B^0}  \rightarrow D^+ D_s^-$
through the spectator mechanism, which does not exist for
 decay mode $\overline{B^0} \rightarrow D_s^+ D_s^-$. There are
 similar diagrams with vector mesons.}}
\label{fig:bdd_fact}
\end{center}
\end{figure}

Based on the symmetry of HQEFT, we  obtain the bosonized currents. 
For a decay of the $b \bar{q}$ system (see Fig. ~\ref{fig:bdd_fact})
we have \cite{ahjoe}:
\begin{equation}
 \overline{q_L} \,\gamma^\mu\, Q_{v_b}^{(+)} \; \longrightarrow \;
 \frac{\alpha_H}{2} Tr\left[\xi^{\dagger} \gamma^\alpha L \, H_{b}^{(+)}
 \right] \; ,
\label{J(0)}
\end{equation}
where (up to QCD and $1/m_Q$ corrections\cite{ahjoe})
 $\alpha_H=f_H \sqrt{m_H}$ for $H=B,D$. Further,
 $Q_{v_b}^{(+)}$ is the heavy $b$-quark field, $v_b$ is
its velocity, and $H_{b}^{(+)}$ is the corresponding heavy meson
field.
For the  $W$-boson materializing to a $\overline{D}$, the bosonized current 
$ \overline{q_L} \gamma^\mu\  Q_{{v_{\bar c}}}^{(-)}$ is given by 
 (\ref{J(0)}) but with  $H_{b}^{(+)}$ replaced by 
$H_{\bar{c}}^{(-)}$ representing  the  $\overline{D}$ meson.
 $v_{\bar c}$ is the velocity of the heavy $\bar{c}$ quark.

The  bosonized $b \rightarrow c$ transition current in
Fig.~\ref{fig:bdd_fact} is given by
\begin{equation}
 \overline{Q_{v_b}^{(+)}} \,\gamma^\mu\, L Q_{v_c}^{(+)}\;\longrightarrow
 \; - \zeta(\omega) Tr\left[ \overline{H_c^{(+)}}\gamma^\alpha L
 H_{b}^{(+)} \right] 
\label{Jbc}
\end{equation}
where $\zeta(\omega)$ is the Isgur-Wise function for the $\bar{B}
\rightarrow D$ - transition, and $v_c$ is the velocity of the heavy
$c$-quark.  Furthermore, $\omega \equiv v_b \cdot v_c= v_b \cdot
v_{\bar c} = M_B/(2M_D)$.  

\begin{figure}[t]
\begin{center}
\includegraphics[width=6cm]{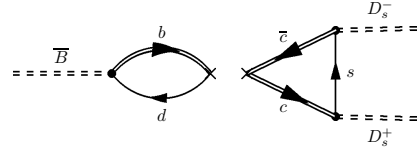} 
\caption{\small{Factorized contribution for 
$\overline{B^0}  \rightarrow D_s^+ D_s^-$
through the annihilation  mechanism, which give zero contributions if
both $D_s^+$ and $D_s^-$ are pseudoscalars.}}
 \label{fig:bdd_fact2}
\end{center}
\end{figure}

For the weak current for $D \overline{D}$ production (corresponding to the
factorizable annihilation mechanism in Fig.~\ref{fig:bdd_fact2}), the current 
$ \overline{Q_{v_c}^{(+)}} \,\gamma^\mu\, L Q_{v_{\bar c}}^{(-)}$
is given by (\ref{Jbc}) with $H_{b}^{(+)}$ replaced by 
$H_{\bar{c}}^{(-)}$, and 
$\zeta(\omega)$ is replaced by $\zeta(-\lambda)$,
where $\lambda= v_{\bar c} \cdot v_c = [M_B^2/(2M_D^2) -1]$.
Note that $\zeta(-\lambda)$ is a complex function which is less known
than $\zeta(\omega)$.

\begin{figure}[t]
\begin{center}
\includegraphics[width=7.5cm]{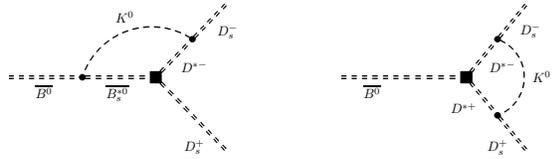}
\caption{Non-factorizable chiral loops for 
$\overline{B^0} \rightarrow D_s^+ D_s^-$. There are similar diagrams
  for vector mesons in the final state.}
\label{fig:chiral1}
\end{center}
\end{figure}

The factorized contributions for the spectator and annihilation
diagrams are shown in Figs. \ref{fig:bdd_fact} ,  \ref{fig:bdd_fact2}.
The first diagram do not give any (direct) contributions to the class of processes
we consider, but is still important because it is the basis of our
chiral loops, visualized in Fig.~\ref{fig:chiral1}. 

The chiral loop amplitudes  are of order $(\ga m_K/4 \pi f)^2$
compared  to  typical factorizable amplitudes in processes where these
exist.

\subsection{The HL$\chi$QM}

The HL$\chi$QM Lagrangian is 
\begin{equation}
{\mathcal L}_{\rm{HL\chi QM}} = {\mathcal L}_{HQEFT} +
{\mathcal L}_{\chi QM} + {\mathcal L}_{Int} \; .
\label{totlag}
\end{equation}
The first term is given in (\ref{LHQEFT}) and the second term 
 is described by the Chiral Quark Model
of the light sector involving 
interactions between quarks and (Goldstone) mesons:
\begin{equation}
{\mathcal L}_{\chi QM} =
\bar \chi \left[\gamma^\mu (i {\cal D}_\mu +   
\gamma_5  {\mathcal A}_{\mu}) - m \right]\chi  \; .
\label{chqmR}
\end{equation}
Here $m=(230\pm20)$MeV is the $SU(3)$ invariant constituent light
quark mass, and $\chi$ is the flavor rotated quark field given by
$\chi_L=\xi^\dagger q_L$ and $\chi_R = \xi q_R$, where $q^T = (u,d,s)$
is the light quark field. 
The covariant derivative ${\cal D}_\mu$  contains the soft
gluon field forming the gluon condensates (besides some chiral interactions) 
\cite{ahjoe,ahjoeB,BEF}.

The interaction between heavy meson fields and quarks is
described by \cite{ahjoe}:
\begin{equation}
{\mathcal L}_{Int}  =   
 -   G_H \, \left[ \chibar_a \, \overline{H_a^{(\pm)}} 
\, Q^{(\pm)}_{v} \,
  +     \overline{Q_{v}^{(\pm)}} \, H_a^{(\pm)} \, \chi_a \right]  
\label{Int}
\end{equation}
where the coupling constant $G_H = \sqrt{2 m
\rho}/f$, 
 and $\rho$ is a hadronic parameter
depending on $m$ (numerically  $\rho$ is of order one \cite{ahjoe})

Performing the bosonization of the HL$\chi$QM, one encounters
divergent loop integrals which will in general be quadratic, linear
and logarithmic divergent \cite{ahjoe}. The quadratic and
logarithmic integrals are related to the
quark condensate and the gluon condensate respectively \cite{ahjoe,BEF}.
The linearly divergent integral is related to $\ga$.

\begin{figure}[t]
\begin{center}
\includegraphics[width=6cm]{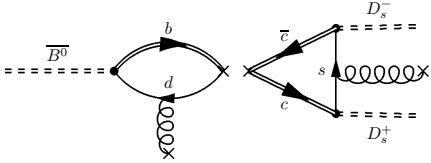}
\caption{Non-factorizable contribution for 
$\overline{B^0}  \rightarrow D_s^+ D_s^-$
through the annihilation mechanism with additional soft gluon emision.
 The wavy lines represent soft
gluons ending in vacuum to make gluon condensates.}
\label{fig:bdd_nfact2}
\end{center}
\end{figure}

The gluon condensate amplitudes can be written, within the
framework presented in the previous section, in a quasi-factorized
way as a product of matrix elements of colored currents:
 as visualized
in Fig. \ref{fig:bdd_nfact2}.  The left part in
Fig. \ref{fig:bdd_nfact2} gives us the bosonized colored current:
\bea
\left(\overline{q_L}\, t^a  \,\gamma^\alpha \, Q_{v_b}^{(+)}\right)_{1G} 
\;   \longrightarrow \; 
- \fr{G_H \, g_s}{64 \pi} \,G_{\mu\nu}^a \nonumber \\
\times Tr\left[\xi^\dagger
\gamma^\alpha L \, H_b^{(+)}
\left( \sigma^{\mu\nu} \, - \, F \,  \{\sigma^{\mu\nu},
 \gamma \cdot v_b \} \, \right)\right] 
\label{1G}
\eea
where $G^a_{\mu \nu}$ is the octet gluon tensor, and
$F \; \equiv \; 2 \pi f^2/(m^2\,N_c)$
is a dimensionless quantity of the order 1/3. The symbol $\{\; , \; \}$
denotes the anti-commutator.
For the creation of a $D \overline{D}$ pair in the right part of Fig.
\ref{fig:bdd_nfact2} the bosonization of the colored current 
$ \left(\overline{Q_{v_c}^{(+)}} \,t^a \; 
\gamma^\alpha \, L Q_{\bar{v}}^{(-)}\right)_{1G} \;$
is bosonized similarly to (\ref{1G}), but involves 
$H_{c}^{(+)}$ and $H_{\bar{c}}^{(-)}$.

Multiplying the two colored  currents
and using   the replacement:
\begin{equation}
g_s^2 G_{\mu \nu}^a G_{\alpha \beta}^a  \; \rightarrow 4 \pi^2
 \gc \frac{1}{12} (g_{\mu \alpha} g_{\nu \beta} -  
g_{\mu \beta} g_{\nu \alpha} ) 
\label{gluecond}
\end{equation}
we obtain a bosonized effective Lagrangian which is $1/N_c$
suppressed compared to the factorized contributions. 
This effective Lagrangian correspond to a certain linear combination
of  a priori possible  $1/N_c$ suppressed terms   at tree level (in
the chiral perturbation theory sense).

\section{RESULTS}

Summing the chiral loops, the gluon condensate contributions and the
annihilation contributions (for vectors in the final state), we obtain,
for chiral loops regularized in the MS-bar scheme,
the branching ratios:  $BR (\bar B^0 \to D_s^+ D_s^-) \simeq 4.1 \times 10^{-4}$, 
  $BR (\bar B^0 \to D_s^{+*} D_s^-)  \simeq 2.8 \times 10^{-4}$,
 $BR (\bar B^0 \to D_s^+ D_s^{-*}) \simeq 3.2 \times 10^{-4}$,
and $BR (\bar B^0 \to D_s^{+*} D_s^{-*}) \simeq (20-50) \times
10^{-4}$. The  masses of $D_s$ and $D_s^*$ are
 equal in the formal $m_c \rightarrow \infty$ limit. Especially the
last branching ratio are sensitive to $1/m_c$ corrections, which
 should be included in a future 
 step of the calculations, like  in \cite{ahjoeB}. The numbers above
 should therefore be considered as preliminary estimates.
For further details, see \cite{EFP}, where results for 
$\overline{B^0_s} \rightarrow D^{(*)} \overline{D^{(*)}}$ 
will  also be presented.

\end{document}